\documentclass[conference]{IEEEtran}
\IEEEoverridecommandlockouts
\usepackage{cite}
\usepackage{amsmath,amssymb,amsfonts}
\usepackage{algorithmic}
\usepackage{graphicx}
\usepackage{textcomp}
\usepackage{xcolor}
\usepackage{booktabs}
\usepackage{booktabs}
\usepackage{floatrow}
\usepackage{siunitx}
\usepackage{adjustbox}
\usepackage{soul}
\usepackage{hyperref}
\usepackage{adjustbox}
\usepackage{comment}

\usepackage[flushleft]{threeparttable}
\usepackage{float}
\floatstyle{plaintop}
\restylefloat{table}
\def\BibTeX{{\rm B\kern-.05em{\sc i\kern-.025em b}\kern-.08em
    T\kern-.1667em\lower.7ex\hbox{E}\kern-.125emX}}
\begin{document}

\title{Effect of Auditory Stimuli on Electroencephalography-based Authentication
}

\author{
\IEEEauthorblockN{
Nibras ABO ALZAHAB\textsuperscript{1},
Angelo DI IORIO\textsuperscript{1},
Marco BALDI\textsuperscript{1}, 
Lorenzo SCALISE\textsuperscript{2}}
\IEEEauthorblockA{
\textit{Università Politecnica delle Marche, Ancona, Italy}\\
\textsuperscript{1}Dept. Information Enginering (DII) \\
\textsuperscript{2}Dept. Industrial Engineering and Mathematical Science (DIISM) \\
n.abo\_alzahab@pm.univpm.it, angelo.di-iorio94@hotmail.it, m.baldi@univpm.it, l.scalise@univpm.it
}
}

\maketitle

\begin{abstract}
Opposed to standard authentication methods based on credentials, biometric-based authentication has lately emerged as a viable paradigm for attaining rapid and secure authentication of users. Among the numerous categories of biometric traits, electroencephalogram (EEG)-based biometrics is recognized as a promising method owing to its unique characteristics.
This paper provides an experimental evaluation of the effect of auditory stimuli (AS) on EEG-based biometrics by studying the following features: i) general change in AS-aided EEG-based biometric authentication in comparison with non-AS-aided EEG-based biometric authentication, ii) role of the language of the AS and ii) influence of the conduction method of the AS.
Our results show that the presence of an AS can improve authentication performance by 9.27\%. Additionally, the performance achieved with an in-ear AS is better than that obtained using a bone-conducting AS. Finally, we verify that performance is independent of the language of the AS. The results of this work provide a step forward towards designing a robust EEG-based authentication system.

\end{abstract}

\begin{IEEEkeywords}
Authentication, biometrics, electroencephalography, machine learning, neural networks.
\end{IEEEkeywords}

\section{Introduction}
The need for cybersecurity and data protection is increasing due to technological improvements in many aspects of everyday life, and user authentication is a first and crucial step in the cybersecurity chain. Authentication involves the usage of credentials, that could be expressed as ``something you have" (possession), ``something you know" (knowledge) or ``something you are" (biometrics) \cite{bidgoly2020survey}. 
The usage of credentials based on the first two techniques (either tokens or passwords) is inherently exposed to the risk of theft and loss of the credentials.
Techniques based on biometrics mitigate these risks and provide a legitimate alternative to owned or known credentials. 
Biometric data cannot be lost or forgotten since they are naturally with the owner, such as fingerprints and iris print \cite{fairhurst2018biometrics}, and are also consequently difficult to be copied or stolen. 
Biometric credentials are usually defined by means of seven terms \cite{jain2011introduction}: universality, uniqueness, permanence, collectability, performance, acceptability and circumvention, as defined next. 
\begin{enumerate}
    \item \textbf{Universality:} It means that every individual should own the biometric data. This guarantees that the biometric data might be utilised by most people.
    \item \textbf{Uniqueness:} It is the most significant factor for identification and indicates that the biometric traits cannot be shared by two or more persons.
    \item \textbf{Permanence:} It refers to the steadiness over time. The biometric data cannot be modified from time to time.
    \item \textbf{Collectability:} Defines how straightforward is to measure the biometric parameters in a quantitative manner.
    \item \textbf{Performance:} Reflects how efficient is to execute identification based on biometrics in terms of accuracy and complexity.
    \item \textbf{Acceptability:} Reflects how individuals are willing to use biometrics in practice and how happy they are with the system.
    \item \textbf {Circumvention:} It is linked to spoofing resistance. In other words, some biometric features could be mimicked easily while for others it is considerably harder.
\end{enumerate}

Traditional biometric features like fingerprints, iris recognition, and signatures can be copied or extracted from a corpse. In contrast to these methods, brain waves provide more difficult-to-forge biometric signals. 
For this reason, systems based on electroencephalography (EEG) have already been considered in the field of cybersecurity, as a basis for authentication \cite{altahat2017robust,goudiaby2020eeg}. 
The main advantage of EEG-based authentication is that EEG signals are generated exclusively by living beings and are mood-dependent.
As a result, they cannot be extracted from a dead brain or through force or threat, which makes them more robust than other biometric signals \cite{bidgoly2020survey}. 
Nevertheless, there are still many challenges to be faced in order to obtain practical EEG-based authentication.

\begin{figure}[ht]
\includegraphics[width=0.7\textwidth]{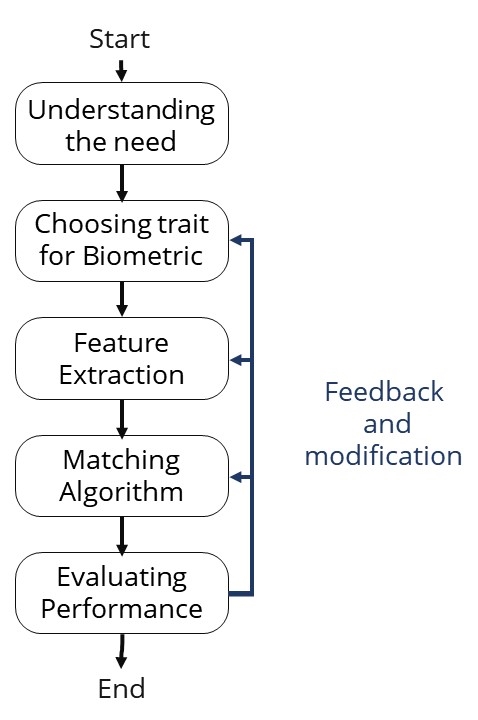}
\caption{Design Cycle of Biometric Systems \cite{alzahab2021efficient}}
\label{Biometric cycle}
\end{figure}

Designing secure and efficient EEG-based authentication systems basically is an open challenge, and the design process must follow the steps presented in Fig.~\ref{Biometric cycle}, as described in \cite{jain2011introduction}:
\begin{enumerate}
    \item Definition of the requirements of the information system and the accompanying security requirements.
    \item Choice of the biometric features that include sufficient identity information, especially those covering the uniqueness requirement.
    \item Data collection and feature extraction. For better outcomes, the collected dataset should involve enough participants to guarantee universality. Since gathered data often contain far more information than needed, feature extraction is required to focus on identity-related information and reject other forms of information that could be deceptive.
    \item Classification through matching algorithms based on the comparison with a template, employing a mathematical model or artificial intelligence and machine learning methods.
    \item Assessment of authentication performance to improve the design with the purpose of obtaining some desired performance target in terms of authentication accuracy.
\end{enumerate}

In \cite{li2020music,li2020authentication} the use of a brain-computer interface with authentication systems in the presence of a musical stimuli has been studied. The results showed that auditory-evoked response carries subject discriminating features, which can be potentially used as a biometric. Another effort in this field was conducted in \cite{marinocognitive}, where the participant in EEG-authentication was exposed to three different genres of music. The experiments were repeated over 6 weeks. The results showed that the reaction of the brain is different when exposed to a familiar music genre. This means that the brain develops specific features after repetition, regardless of the genres of the music. 
Based on these premises, it seems legitimate to investigate whether the language has a similar effect or not.
Moreover, auditory stimuli can be conveyed through in-ear or bone-conducting headphones, thus one may also wonder if the conduction method affects the performance of EEG-based authentication.
The aim of this paper is to study the effect of the auditory stimuli on the performance of EEG-based biometric authentication, focusing on these aspects.
In particular, we aim at identifying whether there is an effect of auditory stimuli on the performance and practical feasibility of EEG-based authentication.
Additionally, this work investigates the effect of language of the auditory stimuli on the authentication performance.

\section{Materials and Methodology}

In this section we describe the datasets we used for assessing the performance of EEG-based authentication along with the chosen authentication methods.

\subsection{Local dataset}

\label{Local Dataset}

A local dataset, which has been made publicly available on Physionet \cite{PhysioNet,alzahabauditory} was recorded at Marche Polytechnic University by enrolling 20 participants who performed the following 8 experiments:

\begin{enumerate}
    \item Three minutes of resting-state, eyes open for three sessions.
    \item Three minutes of resting-state, eyes closed for three sessions.
    \item Recording EEG signal while hearing a song in the native language using in-ear headphone.
    \item Recording EEG signal while hearing a non-native language song using in-ear headphone.
    \item Recording EEG signal while hearing neutral music using in-ear headphone.
    \item Recording EEG signal while hearing a song in your native language using bone-conducting headphone.
    \item Recording EEG signal while hearing a non-native language song using bone-conducting headphone.
    \item Recording EEG signal while hearing neutral music using bone-conducting headphone.
\end{enumerate}
The EEG signals were captured from four channels, namely T7, F8, Cz, and P4, at a sampling rate of 200 Hz. Data preprocessing comprises a first-order bandpass Butterworth filter with a frequency range of 3 - 40 Hz.
To ascertain the subjects' level of comfort during the recording, they were asked to rank the experiments in order of their satisfaction. For the sake of simplification, the experiments were divided into four categories:
\begin{enumerate}
    \item Resting-state: Eyes Open.
    \item Resting-state: Eyes Close.
    \item Auditory Stimuli using in-ear headphone.
    \item Auditory stimuli using bone-conducting headphone.
\end{enumerate}

\subsection{Auditory stimuli analysis}
The purpose of this analysis is to provide answers to the following three questions:
\begin{enumerate}
    \item Do auditory stimuli affect the performance of EEG-based biometric authentication?
    \item Dose the auditory conduction method affect the performance of EEG-based biometric authentication?
    \item Does the EEG-based biometric authentication performance differ between native, non-native, and neutral music?
\end{enumerate}

\begin{table*}[ht!]
    \caption{MLP Architectures}
    \centering
    \begin{adjustbox}{max width = \columnwidth}
    \begin{tabular}{ccccc}
    \toprule[1.5pt]
         & \multicolumn{2}{|c|}{\textbf{\textit{STEW (Resting State \& Mental Load)}}} & \multicolumn{2}{c|}{\textbf{\textit{EEG Alpha Wave dataset}}}  \\
         Layer & Number of Neurons & Activation Function & Number of Neurons & Activation Function \\
         
         \hline 
         Dense 1    &   200 &   ReLU    &   200 & ReLU  \\
         Dense 2    &   150 &   ReLU    &   120 & ReLU  \\
         Dense 3    &   100 &   ReLU    &   70 & ReLU   \\
         Dense 4    &   75 &   ReLU    &   19 & Softmax  \\
         Dense 5    &   48 &   Softmax    & & \\
         
         \bottomrule[1.5pt]
    \end{tabular}
    \end{adjustbox}
    \label{architecture}
\end{table*}
\begin{figure*}
    \centering
    \includegraphics[width=\textwidth]{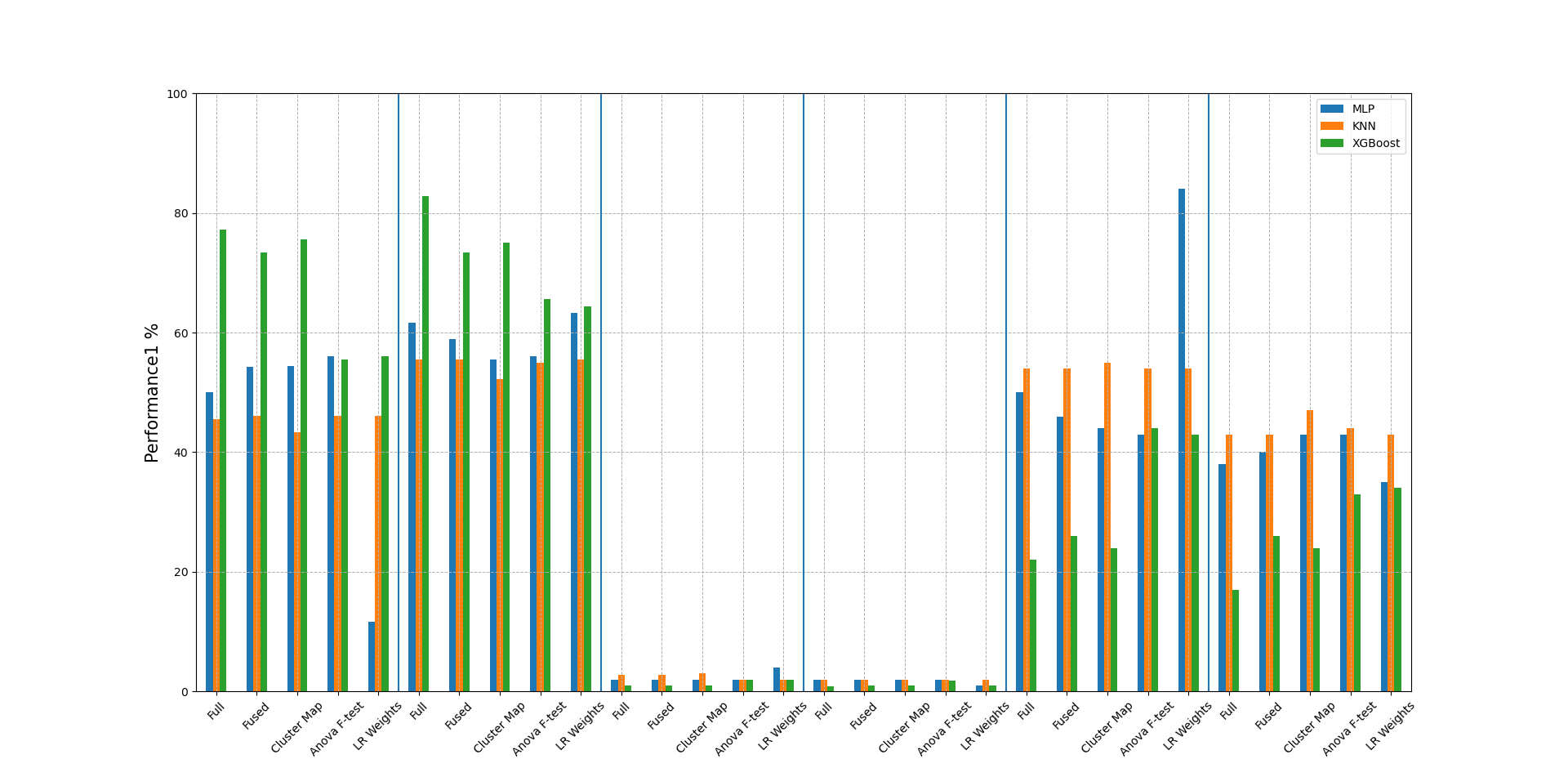}
    \caption{Authentication performance (Accuracy, FAR, and FRR): Ex02 (Resting State Closed eyes) versus Ex07 (Auditory Stimuli)}
    \label{Local_AS}
\end{figure*}

These questions were addressed by using the locally recorded EEG dataset. To do this, we considered eight EEG-based biometric authentication systems based on the studies performed to collect the dataset.
Following the EEG duration experiment conducted in \cite{abo_alzahab_2021}, the EEG-epoch was split into 4-second segments; By EEG-epoch we mean a specific time-windows extracted from the continuous EEG signal.
Instead, the features were extracted from the cluster map dataset in accordance with the results of a previous finding \cite{alzahab2021efficient}. Three distinct classifiers were employed for classification: Multilayer Perceptron (MLP), K-Nearest Neighbors (KNN), and eXtream Gradient Boosting (XGBoost).
In order to assess the user satisfaction for each method, the subjects were asked to rank the four experiments' types:  In-Ear Auditory Stimuli, and Bone-Conducting Auditory Stimuli, Eyes Open Resting State, Eyes Closed Resting State.

\subsection{Authentication methods}

In this section we describe the tools used for performing authentication based on the considered EEG signals and assess the corresponding performance.

\subsubsection{Multilayer perceptron (MLP)}

Each dataset was classified using a unique neural network architecture created using a trial and error approach. Five dense layers comprise the architecture utilized to classify the STEW data. Each layer has 200, 150, 100, 57, and 48 neurons, respectively. On the other hand, the classification architecture for the EEG-alpha dataset consists of the following four layers: 200, 120, 70, and 19 neurons. Cross-entropy was employed as the loss function and the Adam optimizer was used in both architectures. Additionally, a batch size of 16 was chosen for training, and 1000 epochs were used. The MLP is summarized in Table ~\ref{architecture}.

\subsubsection{k Nearest Neighbours (KNN)}

KNN is a simple classifier that uses majority votes to determine class membership. The vote is limited to a specified number of nearest neighbours. We evaluated a range of neighbours from 1 to 20, with the greatest performance obtained when $K = 1$.
\subsubsection{eXtream Gradient Boosting (XGBoost)}
XGBosst is a cutting-edge classifier that was introduced in \cite{chen2016xgboost}.
It is an ensemble method that enhances the performance of simpler models by combining them together. 
It is regarded as a promising method due to its great performance and short computing time \cite{ogunleye2019xgboost}.

\begin{figure}[ht]
    \centering
    \includegraphics[width=\textwidth]{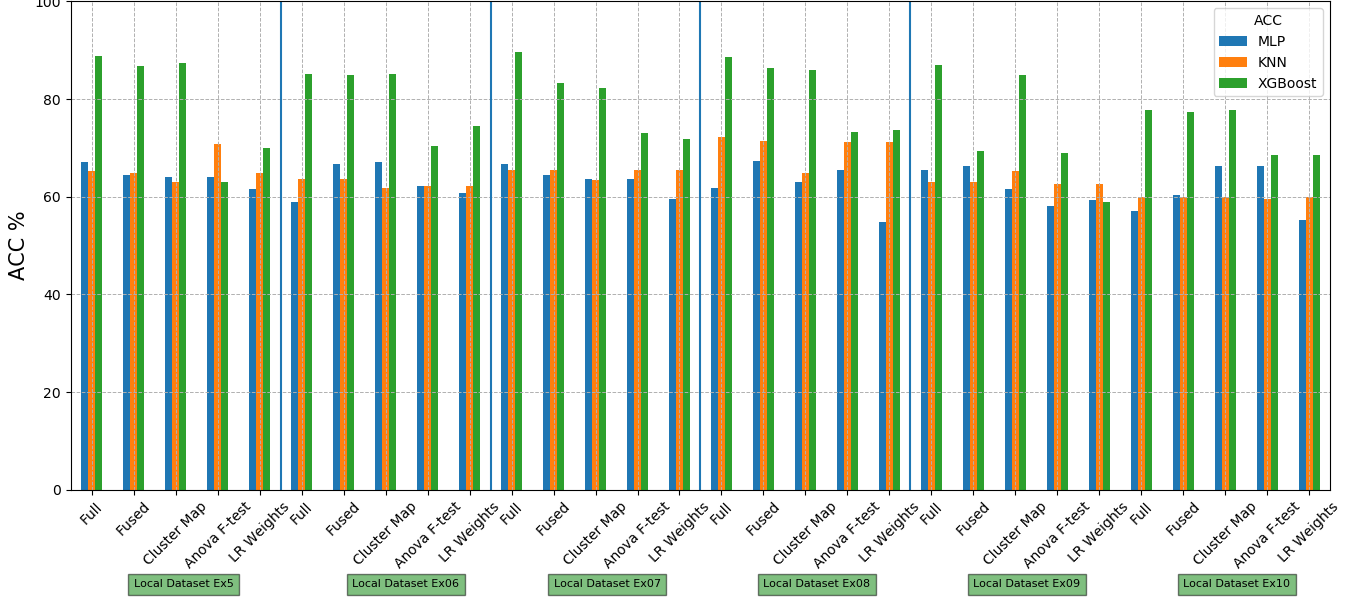}
    \caption{Authentication Performance with different auditory stimuli (Accuracy).}
    \label{Local_ACC}
\end{figure}

\begin{figure}[ht]
    \centering
    \includegraphics[width=\textwidth]{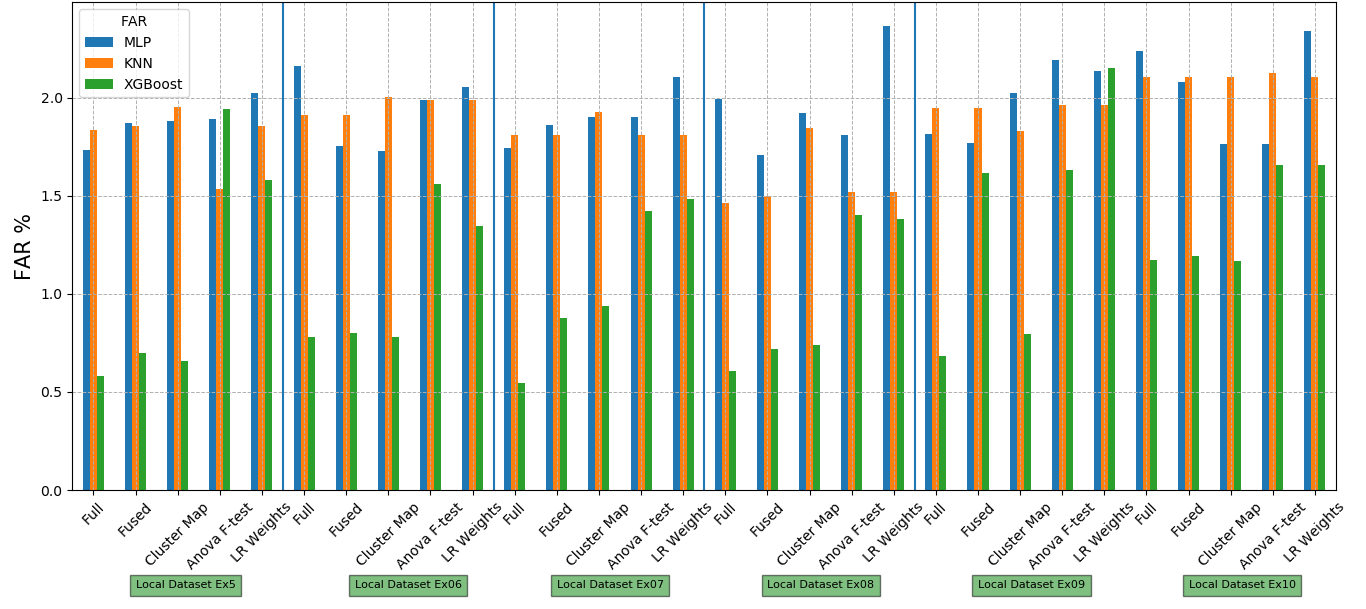}
    \caption{Authentication Performance with different auditory stimuli (FAR).}
    \label{Local_FAR}
\end{figure}

\begin{figure}[ht]
    \centering
    \includegraphics[width=\textwidth]{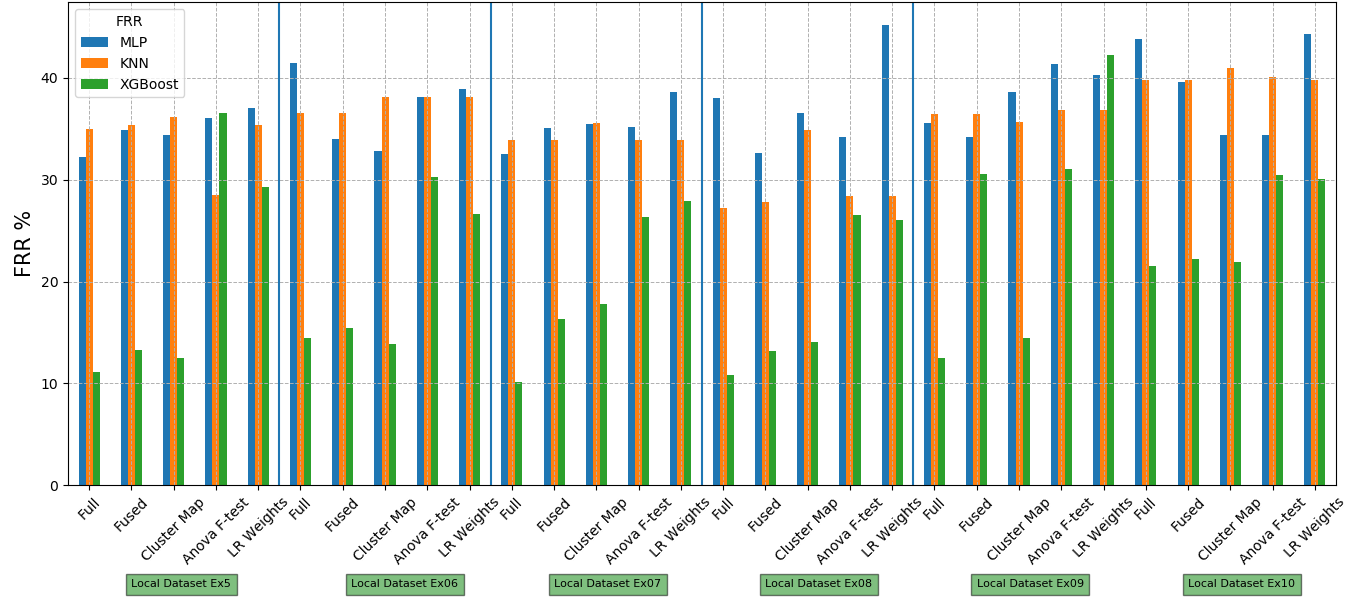}
    \caption{Authentication Performance with different auditory stimuli (FRR).}
    \label{Local_FRR}
\end{figure}

For all the considered methods, the classification process was repeated three times to increase robustness of the results against statistical oscillations. We then computed the average and standard deviation of performance measures.

\section{Results}
\begin{figure}
    \centering
    \includegraphics[width=\textwidth]{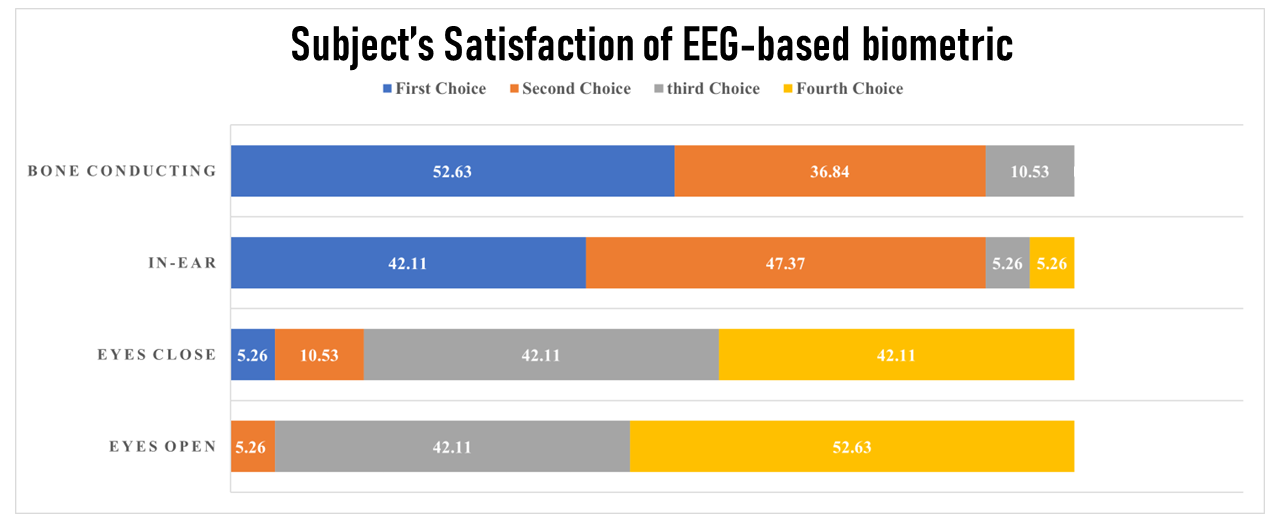}
    \caption{Subject's Satisfaction of EEG-based biometric authentication}
    \label{Subject_Satisfication}
\end{figure}

To assess the influence of the auditory stimuli, the local dataset was utilized to combine the results of the prior two trials. The results of a study to determine if auditory stimuli impact EEG-based biometric authentication performance are reported in Fig. \ref{Local_AS}. It compares the accuracy, FAR, and FRR of Experiment 02 (Resting-State: Eyes Closed) and Experiment 07 (Resting-State: Auditory Stimuli Hearing Neutral Music Using In-Ear Headphones). It demonstrates a 9.27\% percent increase in accuracy when auditory stimuli are used.
In order to assess the influence of the sound conducting method and the language of the auditory stimuli, in Figs. ~\ref{Local_ACC}, ~\ref{Local_FAR}, and ~\ref{Local_FRR} we report the accuracy, FAR, and FRR of three classifiers: MLP, KNN, and XGBoost. The results indicate that there is no statistically significant difference between the two cases. Additionally, Fig.~\ref{Subject_Satisfication} depicts the subjects' satisfaction as determined by the survey introduced in Section~\ref{Local Dataset}. Our findings indicate that using auditory stimuli is preferable to the resting state condition. Additionally, bone-conducting treatments are more gratifying than in-ear stimulation.

\section{Discussion}

Based on the previously described experiments and the relevant results, we can make the following observations.
\paragraph{Effect of the auditory stimuli on performance of EEG-based biometric authentication} 
    As introduced in Section \ref{Local Dataset}, Fig.~\ref{Local_AS} compares Ex02 and Ex07, which were conducted in the resting state and with auditory stimuli, respectively. Notably, the auditory stimuli instance exceeded the resting condition by a difference of 9.27\%  in accuracy. Such a result comes inline with the findings of \cite{li2020music,li2020authentication}. This can be explained by the fact that the brain's response to auditory stimuli generates distinct EEG oscillation patterns. As a result, the addition of auditory stimuli can improve authentication performance.
    In terms of implementability, as illustrated in Fig.~\ref{Subject_Satisfication}, 52.63\%  of subjects preferred auditory stimuli produced using bone-conducting headphones as their first choice for satisfaction, followed by 36.84\%  as a second choice. While 42.11\%  of subjects chose auditory stimuli via in-ear headphones as their primary source of satisfaction, 47.37\%  chose it as a secondary source of satisfaction. This suggests that auditory stimuli are preferable in terms of performance and practical feasibility. 
    
    \paragraph{Effect of the auditory conduction method on the performance of EEG-based biometric authentication}
    The paired t.test was used to compare bone-conducting stimuli to in-ear auditory stimuli. P=0.0380.05 for accuracy, P=0.0460.05 for FAR, and P=0.0320.05 for FRR are the test results. As a result, there is a significant difference between auditory stimuli sent through the ear canal and auditory stimuli delivered through the bone. The average accuracy of the in-ear case (69.33 ± 8.92\%) is slightly greater than that of the bone-conducting case (67.60 ± 8.78\%).
    In terms of user satisfaction and implementability, according to the survey results shown in Fig.~\ref{Subject_Satisfication}, 42.11\%  of subjects preferred in-ear auditory stimuli, whereas 52.63\%  preferred bone-conducting auditory stimuli.
    This trade-off between performance and implementability allows the system designer to prioritize either one or the other. 
    
    \paragraph{Differences in the EEG-based biometric authentication performance between native, non-native, and neutral music}
     The ANOVA test was used to determine whether there is a significant difference between the three groups. The statistical analysis produced a p-Value $> 0.05$, indicating that the difference is not significant. There is no correlation between EEG-based biometric authentication performance and the auditory stimuli's language. This is consistent with the findings in \cite{marinocognitive}, where it was concluded that EEG-based biometric authentication is genre-independent.
    
\section{Conclusion}
In conclusion, our experiments confirm that the use of EEG-based biometric authentication has the potential to represent a new cybersecurity tool with unique features.
This work contributes to the study of the performance achievable by EEG-based biometric authentication with the following summary results:
\begin{enumerate}
    \item Using auditory stimuli could improve the authentication performance by more than 9\%.
    \item Using in-ear auditory stimuli is better than using bone-conducting auditory stimuli in terms of performance, despite bone-conduction turns out to be more acceptable by users than in-ear conduction.
    \item Performance of EEG-based biometric authentication in the presence of an auditory stimulus is independent of the language of the auditory stimulus.
\end{enumerate}

\bibliographystyle{ieeetr}
\bibliography{Biblography.bib}

\begin{thebibliography}{10}

\bibitem{bidgoly2020survey}
A.~J. Bidgoly, H.~J. Bidgoly, and Z.~Arezoumand, ``A survey on methods and
  challenges in eeg based authentication,'' {\em Computers \& Security},
  p.~101788, 2020.

\bibitem{fairhurst2018biometrics}
M.~Fairhurst, {\em Biometrics: A Very Short Introduction}.
\newblock Oxford University Press, USA, 2018.

\bibitem{jain2011introduction}
A.~K. Jain, A.~A. Ross, and K.~Nandakumar, {\em Introduction to biometrics}.
\newblock Springer Science \& Business Media, 2011.

\bibitem{altahat2017robust}
S.~H.~Q. Altahat, {\em Robust {EEG} Channel Set for {B}iometric Application}.
\newblock PhD thesis, University of Canberra, 2017.

\bibitem{goudiaby2020eeg}
B.~Goudiaby, A.~Othmani, and A.~Nait-Ali, ``Eeg biometrics for person
  verification,'' in {\em Hidden Biometrics}, pp.~45--69, Springer, 2020.

\bibitem{alzahab2021efficient}
N.~Abo~Alzahab, M.~Baldi, and L.~Scalise, ``Efficient feature selection for
  electroencephalogram-based authentication,'' in {\em 2021 IEEE International
  Symposium on Medical Measurements and Applications (MeMeA)}, pp.~1--6, IEEE,
  2021.

\bibitem{li2020music}
S.~Li, L.~Marino, and V.~Alluri, ``Music stimuli for eeg-based user
  authentication,'' in {\em The Thirty-Third International Flairs Conference},
  2020.

\bibitem{li2020authentication}
S.~Li and M.~Qiu, ``Authentication study for brain-based computer interfaces
  using music stimulations,'' in {\em International Conference on Algorithms
  and Architectures for Parallel Processing}, pp.~663--675, Springer, 2020.

\bibitem{marinocognitive}
L.~Marino, V.~Alluri, A.~Kumar, S.~Li, A.~Leider, and C.~Tappert, ``Cognitive
  biometrics,'' in {\em Student-Faculty Research Day}, pp.~1--5, CSIS, Pace
  Univ., 2020.

\bibitem{PhysioNet}
A.~L. Goldberger, L.~A.~N. Amaral, L.~Glass, J.~M. Hausdorff, P.~C. Ivanov,
  R.~G. Mark, J.~E. Mietus, G.~B. Moody, C.-K. Peng, and H.~E. Stanley,
  ``{PhysioBank, PhysioToolkit, and PhysioNet}: Components of a new research
  resource for complex physiologic signals,'' {\em Circulation}, vol.~101,
  no.~23, pp.~e215--e220, 2000 (June 13).
\newblock Circulation Electronic Pages:
  http://circ.ahajournals.org/content/101/23/e215.full PMID:1085218; doi:
  10.1161/01.CIR.101.23.e215.

\bibitem{alzahabauditory}
N.~Abo~Alzahab, A.~Di~Iorio, L.~Apollonio, M.~Alshalak, A.~Gravina,
  L.~Antognoli, M.~Baldi, L.~Scalise, and B.~Alchalabi, ``Auditory evoked
  potential eeg-biometric dataset.''
  \url{https://physionet.org/content/auditory-eeg/1.0.0/}, 2021.
\newblock Accessed: 17/02/2022.

\bibitem{abo_alzahab_2021}
N.~Abo~Alzahab, ``Design and implementation of techniques for the secure
  authentication of users based on electroencephalogram ({EEG}) signals,''
  Master's thesis, Marche Polytechnic University, 2021.

\bibitem{chen2016xgboost}
T.~Chen and C.~Guestrin, ``Xgboost: A scalable tree boosting system,'' in {\em
  Proceedings of the 22nd acm sigkdd international conference on knowledge
  discovery and data mining}, pp.~785--794, 2016.

\bibitem{ogunleye2019xgboost}
A.~Ogunleye and Q.-G. Wang, ``Xgboost model for chronic kidney disease
  diagnosis,'' {\em IEEE/ACM transactions on computational biology and
  bioinformatics}, vol.~17, no.~6, pp.~2131--2140, 2019.

\end{thebibliography}

\end{document}